\documentclass[aps,twocolumn,showpacs,superscriptaddress,floatfix]{revtex4-1}
\usepackage{graphicx}
\usepackage{amssymb}

\begin{document}

\title{Superconductivity and magnetic order in the non-centrosymmetric Half Heusler compound ErPdBi}
\author{Y. Pan} \affiliation{Van der Waals - Zeeman Institute, University of Amsterdam, Science Park 904, 1098 XH Amsterdam, The Netherlands}
\author{A. M. Nikitin} \affiliation{Van der Waals - Zeeman Institute, University of Amsterdam, Science Park 904, 1098 XH Amsterdam, The Netherlands}
\author{T. V. Bay} \affiliation{Van der Waals - Zeeman Institute, University of Amsterdam, Science Park 904, 1098 XH Amsterdam, The Netherlands}
\author{Y. K. Huang} \affiliation{Van der Waals - Zeeman Institute, University of Amsterdam, Science Park 904, 1098 XH Amsterdam, The Netherlands}
\author{C. Paulsen} \affiliation{Institut N\'{e}el, CNRS, and Universit\'{e} Joseph Fourier, BP 166, 38042 Grenoble, France}
\author{B. H. Yan} \affiliation{Max-Planck-Institut f\"{u}r Chemische Physik fester Stoffe, N\"{o}thnitzer Strasse 40, 01187 Dresden, Germany}
\author{A. de Visser} \email{a.devisser@uva.nl} \affiliation{Van der Waals - Zeeman Institute, University of Amsterdam, Science Park 904, 1098 XH Amsterdam, The Netherlands}

\date{\today}

\begin{abstract}We report superconductivity at $T_c = 1.22$~K and magnetic order at $T_N = 1.06$~K in the semi-metallic noncentrosymmetric Half Heusler compound ErPdBi. The upper critical field, $B_{c2}$, has an unusual quasi-linear temperature variation and reaches a value of 1.6~T for $T \rightarrow 0$. Magnetic order is found below $T_c$ and is suppressed at $B{_M} \sim 2.5$~T for $T \rightarrow 0$. Since $T_c \simeq T_N$, the interaction of superconductivity and magnetism is expected to give rise to a complex ground state. Moreover, electronic structure calculations show ErPdBi has a topologically nontrivial band inversion and thus may serve as a new platform to study the interplay of topological states, superconductivity and magnetic order.

\end{abstract}


\maketitle

\section{Introduction} The ternary compound ErPdBi belongs to the Rare Earth palladiumbismuthide ($RE$PdBi) series, which is part of the large family of Half Heusler compounds that crystallize in a cubic structure with 1:1:1 composition. Half Heusler compounds  attract ample attention as multifunctional materials in the fields of spintronics and thermoelectricity, but also as tunable laboratory tools to study a wide range of intriguing physical phenomena, such as half metallic magnetism, giant magnetoresistance and Kondo and heavy fermion physics~\cite{Graf2011}. More recently, a strong interest in Half Heusler compounds with significant spin-orbit coupling has been generated by first-principle calculations~\cite{Chadov2010,Lin2010,Feng2010} that predict an inverted band order, which may give rise to topological quantum states because of the non-trivial $Z_2$ topology~\cite{Hasan&Kane2010,Qi&Zhang2010}. Prominent candidate materials are the $T$PtBi and $T$PdBi series, where $T$ is Y or Sc or a non-magnetic RE element. Interestingly, some of the platinumbismuthides that exhibit band inversion have been reported to superconduct, which makes them promising candidates for topological superconductivity: LaPtBi ($T_c = 0.9$~K~\cite{Goll2008}), YPtBi ($T_c = 0.77$~K~\cite{Butch2011,Bay2012b}) and LuPtBi ($T_c = 1.0$~K~\cite{Tafti2013}). Moreover, since the crystal structure lacks inversion symmetry, unconventional Cooper pair states, notably mixed even and odd parity states, are predicted to make up the superconducting condensate~\cite{Frigeri2004}. This provides a strong motivation to search for similar phenomena in the palladiumbismuthides.

The REPdBi compounds crystallize, just like the REPtBi series, in the cubic structure with the non-centrosymmetric  $F\overline{4}3m$ space group~\cite{Haase2002}. The magnetic and transport properties of the REPdBi series (RE= Er, Ho, Dy, Gd and Nd) were first reported in Refs.~\cite{Riedemann1996,Gofryk2005,Gofryk2011}. Susceptibility data, taken on arc-melted polycrystalline samples, showed antiferromagnetic order for the Ho, Dy, Gd and Nd compounds with N\'{e}el temperatures, $T_N$, of 2, 3.5, 13 and 4.2 K, respectively. ErPdBi did not show magnetic order down to the lowest temperature measured, $T= 1.7$~K. The susceptibility, $\chi (T)$, of ErPdBi follows the Curie-Weiss law with an effective moment $\mu_{eff} = 9.2~\mu_B$, close to the Er$^{3+}$ free ion value of 9.58~$\mu_B$, and a paramagnetic Curie temperature $\Theta_P = -4.6$~K~\cite{Riedemann1996,Gofryk2005}. Transport measurements revealed a semi-metallic-like behaviour with a carrier density $n(4$~K$) = 6.7 \times 10^{19}$~cm$^{-3}$. ErPdBi received furthermore interest because of its thermoelectric effects~\cite{Kaczorowski2005,Sekimoto2006}.

Here we report electrical transport, ac-susceptibility and dc-magnetization measurements on ErPdBi single crystals that provide solid evidence for superconductivity at 1.22~K and magnetic order at 1.06~K. The combination of superconductivity and magnetic order is unusual. Moreover, electronic structure calculations show ErPdBi has an inverted band order and thus should harbor topological quantum states.

\section{Experimental} A single crystalline batch of ErPdBi was prepared out of Bi flux. As starting materials served the elements Er, Pd and Bi with a purity of 3N5, 4N and 5N, respectively. An ingot of ErPdBi was prepared by arc-melting and placed in an alumina crucible with excess Bi flux. The crucible and contents were heated in a quartz tube under a pressure of 0.3 bar high-purity Argon gas to 1150 $^{\circ}$C and kept at this temperature for 36 h. Then the tube was slowly cooled to 500~$^{\circ}$C at a rate of 3~$^{\circ}$C per hour to form the crystals. Scanning Electron Microscopy and Electron Probe Micro Analysis confirmed the main phase is ErPdBi with composition 1:1:1. Bi precipitates are found in the form of thin lines on the surface of the crystals. Powder X-ray diffraction confirmed the $F\overline{4}3m$ space group and the extracted lattice parameter, $a= 6.595$~{\AA}, is in perfect accord with the literature~\cite{Haase2002}. The Bi precipitates give rise to additional tiny peaks in the diffraction patterns. From their intensity we estimate a Bi volume fraction of about 4\%. Single crystals, with typical dimensions $3 \times 2 \times 0.3$~mm$^3$, were carefully cut from the ingot by spark erosion thereby avoiding the Bi precipitates. Their single-crystalline nature was checked by Laue backscattering. After cutting, the surface of the samples was cleaned by polishing. Magnetic characterization in the temperature interval 1.8-300~K was made in a Physical Property Measurements System (Quantum Design). The Curie-Weiss behaviour was confirmed and the values $\mu_{eff} = 3.54~\mu_B$ and $\Theta_P = -3.5$~K are close to the ones reported in Refs.~\cite{Riedemann1996,Gofryk2005}. The Hall effect and resistivity were measured using a MaglabExa system (Oxford Instruments) for $T=$ 4 - 300~K. Resistance and ac-susceptibility measurements were made in a $^3$He refrigerator (Heliox, Oxford Instruments) for $T=0.24 - 10$~K using a low-frequency ($f \leq 313$~Hz) lock-in technique and low excitation currents ($I \leq 100 ~\mu$A). Additional low-temperature dc-magnetization and ac-susceptibility measurements were made using a SQUID magnetometer, equipped with a miniature dilution refrigerator, developed at the N\'{e}el Institute.

\section{Results} 
\begin{figure}
\includegraphics[width=8cm]{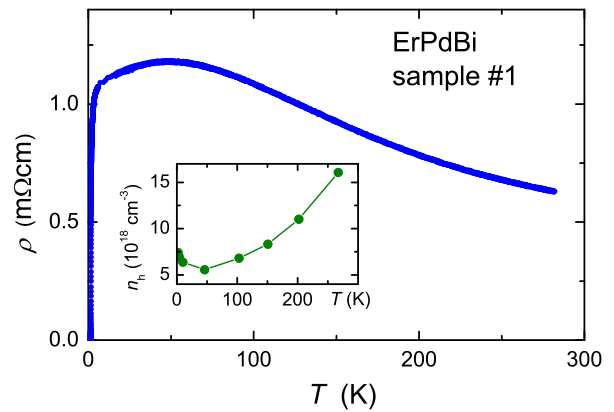}
\caption{(Color on-line) Resistivity and carrier concentration (inset) \textit{versus} temperature of ErPdBi sample \#1.}
\label{Fig.1}
\end{figure}

In Fig.~\ref{Fig.1} we show the resistivity $\rho (T)$ of a flux grown single crystal of ErPdBi (sample \#1). The overall behaviour is in good agreement with the data in the literature with a broad maximum centered at 50 K, rather than 140 K~\cite{Gofryk2011}. The hole carrier concentration, calculated from the low-field linear Hall resistance, equals $n_h = 7.4\times 10^{18}$~cm$^{-3}$ at $T=4$~K (see inset), which is a factor 10 lower than reported in Ref.~\cite{Gofryk2011}. These transport parameters confirm semimetallic-like behaviour. At low temperatures the drop to resistance $R=0$ signals the transition to the superconducting state.

\begin{figure}
\includegraphics[width=8cm]{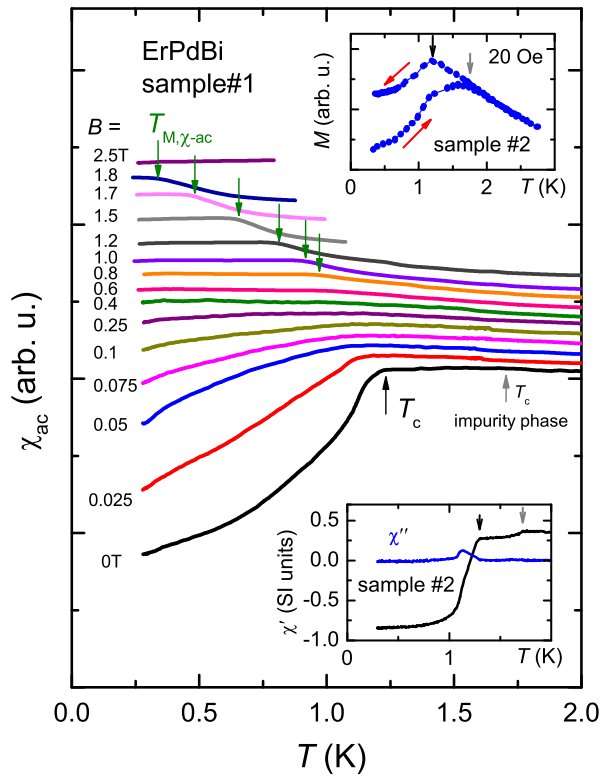}
\caption{(Color on-line) AC susceptibility of ErPdBi (sample \#1) in zero and magnetic fields up to 2.5 T as indicated. The data were taken while cooling in field. Curves are displaced vertically to prevent overlap. The driving field is 0.026~Oe for $B \leq 0.4$~T and 0.26 Oe for $B \geq 0.6$~T. The superconducting transition temperature of ErPdBi and the impurity phase are indicated by black and grey arrows. For $B \geq 0.6$~T the weak local maximum locates the magnetic transition at $T_{M,{\chi_{ac}}}$ (green arrows). Lower inset: $\chi \prime$ and $\chi \prime \prime$ of ErPdBi (sample \#2) in a driving field of 0.027~Oe. Upper inset: Magnetization measured in a field of 20 Oe applied after cooling in zero field; data taken upon warming show screening effects, while data taken upon cooling show flux expulsion (see text).}
\label{Fig.2}
\end{figure}

In Fig.~\ref{Fig.2} we show ac-susceptibility data taken on the same ErPdBi sample that reveal superconductivity occurs below $T_c = 1.22(2)$~K. The superconducting transition appears as a large diamagnetic contribution, which corresponds to a screening fraction of $\sim 92\%$ of the ideal value $\chi_M = -1/(1-N)$ (here $N \simeq 0.1$ is the demagnetization factor). $\chi_{ac}$ data taken on a second sample (\#2) with $N \simeq 0.15$ in a different experimental set-up are reported in the lower inset of Fig.~\ref{Fig.2}. For this sample the screening fraction attains a value of $\sim 90\%$. It should be noted that in both experiments a decrease of the $\chi_{ac}$ signal becomes visible already at a higher temperature, $1.72(2)$~K (as indicated by the grey arrows in Fig.~\ref{Fig.2}). This signal we attribute to an impurity phase with a screening fraction of $\sim 8-10\%$. A large diamagnetic signal is normally a good indicator of bulk superconductivity. Solid proof may be obtained by the observation of flux expulsion. In the upper inset of Fig.~\ref{Fig.2} we show the dc-magnetization measured in a field of 20 Oe after cooling in zero field. Upon warming, the signal is dominated by screening effects due to ErPdBi (up to 1.22~K) and the impurity phase (up to 1.72~K). Upon cooling in field, flux expulsion is predominantly found at $T_c = 1.22$~K. The change in magnetization corresponds to a Meissner fraction of $\sim 15$~\%. This confirms the bulk origin of superconductivity in ErPdBi.

The nature of the impurity phase remains to be solved. Small amounts of Bi inclusions in the form of precipitates might be present in the crystals, but crystalline Bi does not superconduct. Amorphous Bi, \textit{e.g.} prepared as thin film, superconducts at $T_c \sim$~6~K~\cite{Matthias1963}, a temperature much higher than observed here. Among the binary Bi-Pd alloys, the only likely candidate is $\alpha$-Bi$_2$Pd, which is reported to superconduct at 1.7 K~\cite{Matthias1963}. However, if present in our samples, the impurity amount is below the detection limit of the X-ray powder diffraction pattern ($\sim$ 2~\%). We remark that in Refs. \cite{Kaczorowski2005,Gofryk2011} a pronounced drop in the resistivity of \textit{arc-melted} ErPdBi samples was reported at 7~K, together with a field depression that mimics superconductivity. However, no corresponding diamagnetic signal was observed and bulk superconductivity at 7~K was discarded.

$\chi_{ac}$-data taken upon cooling in applied magnetic fields show the diamagnetic screening signal is rapidly lost (see Fig.~\ref{Fig.2}). Surprisingly, in the field range 0.6~T $\leq B < 2.5$~T a pronounced structure appears in $\chi_{ac} (T)$ at temperatures labeled $T_{M,\chi_{ac}}$. Such a (relative) maximum, albeit weak, normally indicates the presence of a magnetic transition. This is corroborated by the field variation of $T_{M,\chi_{ac}}$, which we will discuss after presenting low-temperature resistivity data, $\rho (T)$.

\begin{figure}
\includegraphics[width=8cm]{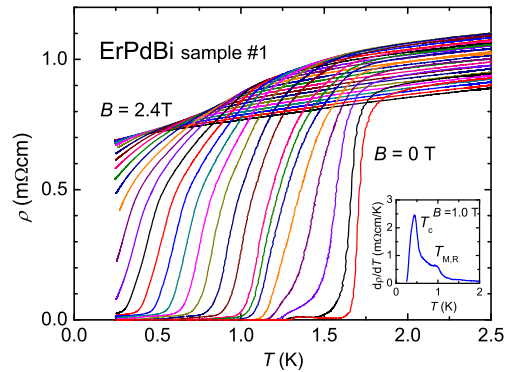}
\caption{(Color on-line) Resistivity of ErPdBi in zero and applied magnetic fields, from right to left: 0, 0.025, 0.05, 0.075, 0.10, 0.13, 0.16, 0.2 T and then up to 2.4 T in steps of 0.1 T. Inset: $d \rho /dT$ \textit{versus} $T$ in $B=1.0$~T. The extrema locate $T_c$ and $T_{M,R}$.}
\label{Fig.3}
\end{figure}

\begin{figure}
\includegraphics[width=8cm]{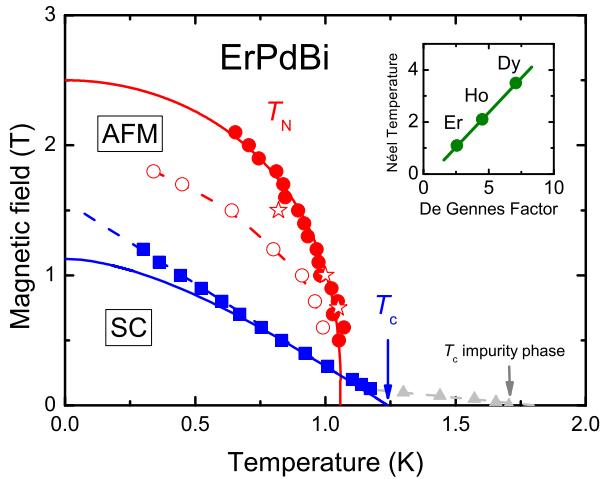}
\caption{(Color on-line) Superconducting (SC) and magnetic (AFM) phase diagram of ErPdBi. Closed blue squares: superconducting transition temperature, $T_c$, determined by extrema in $d \rho /dT$; solid blue line: $B_{c2} (T)$ WHH model curve (see text) with $B_{c2}^{orb}(0) = 1.13$~T. Grey triangles and dashed grey line: $T_c (B)$ of the impurity phase.  Closed circles: $T_N = T_{M,R}$ determined by extrema in $d \rho /dT$; open circles and stars: ($T,B$)$-$location of weak maximum in $\chi_{ac} (T)$ ($T_{M,\chi_{ac}}$) and dc-magnetization (sample \#2); solid red line: magnetic order parameter fit with $T_N =1.06$~K at $B = 0$~T (see text). Inset: De Gennes scaling plot for Er, Ho and DyPdBi (see text).}
\label{Fig.4}
\end{figure}

The superconducting transition in $\rho (T)$ in zero and applied magnetic fields is reported in Fig.~\ref{Fig.3}. In zero field the superconducting transition is due to the impurity phase, where we remark that the transition temperature, $1.74(2)$~K, determined by the midpoint, nicely coincides with the onset temperature, $1.72(2)$ K, in the $\chi_{ac}$-data. In a magnetic field superconductivity of the impurity phase is depressed at the fast rate $dT_c /dB = - 4.4$~K/T (see Fig.~\ref{Fig.4}). Consequently, for $B \geq 0.1$~T and $T \lesssim 1.2$~K the superconducting transition in $\rho (T)$ is due to ErPdBi. At the same time, $\rho (T)$ obtains an unusual round shape around $T_c$. We have determined the upper critical field $B_{c2}$ (or $T_c (B)$) by locating the maximum in $d \rho /dT$ measured at fixed magnetic field, as shown for example for $B=1.0$~T in the inset of Fig.~\ref{Fig.3}. The results are traced in the phase diagram Fig.~\ref{Fig.4}. $B_{c2} (T)$ of ErPdBi displays an unusual quasi-linear temperature variation, which extrapolates to $T_c = 1.24(2)$~K in zero field, close to the onset temperature $T_c = 1.22(2)$~K extracted from $\chi_{ac}$.

The magnetic transition is also detected in the resistance by the local maximum in $d \rho /dT$, as shown in the inset of Fig.~\ref{Fig.3} (temperature labeled $T_{M,R}$). We have traced $T_{M,R}(B)$ and $T_{M,\chi_{ac}}(B)$ in the phase diagram Fig.~\ref{Fig.4}. Both temperatures track the same phase boundary. The location of weak maxima (see Fig.~\ref{Fig.4}) observed in the dc-magnetization (data not shown) for sample \#2 confirm this.

The magnetic transition is almost certainly to an antiferromagnetic (AFM) state with N\'{e}el temperature $T_N$. For $T_N = T_{M,R}$ the phase boundary  obeys the phenomenological order parameter function $B_M (T) = B_M (0)(1-(T/T_N)^{\alpha})^{\beta}$ with $T_N = 1.06$~K, $B_M (0) =2.5$~T, $\alpha = 2$ and $\beta = 0.4$. The latter value is close to the value $\beta = 0.38 $ expected for the 3D Heisenberg antiferromagnet~\cite{Domb1996}. The phase boundaries located by the transport and magnetic data are closely linked, since they all extrapolate to $T_N = 1.06$~K for $B \rightarrow 0$. Local moment AFM order is widely present in the REPdBi series~\cite{Riedemann1996,Gofryk2005}. Strong support for an antiferromagnetic groundstate in ErPdBi is furthermore found in the De Gennes scaling for the heavy rare earth palladium bismuthides (see inset Fig.~\ref{Fig.4}): $T_N \propto (g_J -1)^2 J(J+1)$ with $g_J$ the Land\'{e} factor~(see \textit{e.g.} \cite{Jensen&Mackintosh1991}. Neutron diffraction and/or NMR experiments would be most welcome to investigate the nature of the magnetic order on the microscopic scale.

\section{Discussion} The combination of local-moment antiferromagnetism and superconductivity is unusual. In general local-moment AFM order and superconductivity tend to compete for the ground state. However, in ErPdBi $T_N \simeq T_c$, which tells us both phenomena have similar energy scales. Given the lack of inversion symmetry and the expected unconventional Cooper-pair state~\cite{Bauer&Sigrist2012}, this could give rise to an interesting interplay of superconductivity and magnetism, and a complex ground state. Experimental signatures for this are the unusual rounded shape of the superconducting transition in $\rho (T)$ and the rapid loss of the diamagnetic screening signal in field. Possibly AFM order and superconductivity occupy different sample regions. In order to answer these important questions muon spin relaxation experiments would be very helpful, since these permit one to probe the different volume fractions.

Several other Erbium based antiferromagnetic superconductors have been reported in the literature. In the Chevrel phases ErMo$_6$S$_8$ and ErMo$_6$Se$_8$~\cite{Shelton1983} AFM order and superconductivity compete, while in the borocarbide ErNi$_2$B$_2$C~\cite{Cho1995} and the Heusler phase ErPd$_2$Sn~\cite{Shelton1986} AFM order and superconductivity coexist. Coexistence of superconductivity and AFM order is also found in a number of non-centrosymmetric materials~\cite{Bauer&Sigrist2012}. A prominent example is CePt$_3$Si with $T_c = 0.75$~K and $T_N$ = ~2.2~K~\cite{Bauer2004}. Interestingly, CePdBi, which has the same crystal structure as ErPdBi, also undergoes a magnetic ($T_M = 2$~K) and superconducting transition ($T_c = 1.4$~K)~\cite{Goraus2013}. However, the experiments were carried out on arc-melted polycrystals and the weak diamagnetic screening (8~\% of the sample volume) was hitherto associated with a disordered phase.

The upper critical field of ErPdBi, reported in Fig.~\ref{Fig.4}, shows an unusual linear temperature variation just like for YPtBi~\cite{Bay2012b}, where it was taken as evidence for an odd-parity component in the superconducting order parameter. For ErPdBi, the rounded transitions in $\rho (T)$ and the presence of AFM order, make the determination of $B_{c2}(T)$ difficult. In the limit $T \rightarrow 0$ $B_{c2}$ extrapolates to 1.6~T. Using this value and with help of the relation $B_{c2} = \Phi _0 / 2 \pi \xi ^2$, where $\Phi _0$ is the flux quantum, we calculate a superconducting coherence length $\xi = 14$~nm. Preliminary magnetization measurements show the lower critical field $B_{c1}$ is very small, and a conservative upperbound is $0.0002$~T, which allows an estimation of the Ginzburg-Landau parameter $\kappa = \lambda / \xi $ via the relation $B_{c2}/B_{c1} =2 \kappa ^2 / ln\kappa$, where $\lambda$ is the penetration depth. For $T \rightarrow 0$, we obtain $\kappa \geq 140$ and  $\lambda \geq 448$~nm. In Fig.~\ref{Fig.4} we also compare the measured $B_{c2}(T)$-values  with the model curve for a weak-coupling spin-singlet superconductor in the clean limit with orbital limiting field (Werthamer-Helfand-Hohenberg [WHH] model~\cite{Werthamer1966}). The  zero temperature orbital critical field is given by $B_{c2}^{orb}=0.72 \times T_c ~|dB_{c2}/dT|_{T_{c}}$ and amounts to 1.13~T. Clearly, the $B_{c2}(T)$-values determined from the resistance data exceed the model curve values when $T/T_c \lesssim 0.5$. This is in line with an unconventional Cooper pair state~\cite{Bay2012b}.

\begin{figure}
\includegraphics[width=8cm]{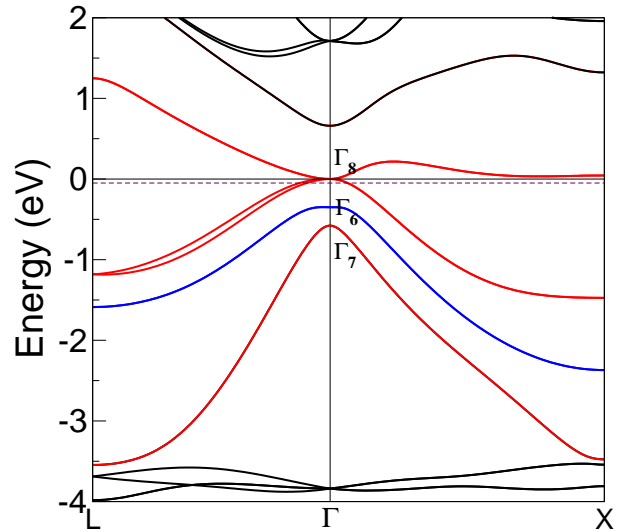}
\caption{(Color on-line) Bulk band structure of half Heusler ErPdBi in the fcc Brillouin zone. The red bands are $\Gamma_8$ and $\Gamma_7$ states and blue is the $\Gamma_6$ state. The Fermi energy is shifted to zero (solid horizontal line). The dashed-horizontal line illustrates the experimental Fermi level with a small hole-pocket at the $\Gamma$ point. Er-$4f$ states were treated as core electrons.}
\label{Fig.5}
\end{figure}

\section{Electronic structure} In order to understand the electronic properties of ErPdBi, we performed \textit{ab initio} band structure calculations based on the density-functional theory within the generalized gradient approximation ~\cite{Perdew1996}. We adopted the Half Heusler structure as determined by experiments. The magnetic susceptibility that follows a Curie-Weiss law~\cite{Riedemann1996,Gofryk2005} reveals the Er-$4f$ electrons are well localized and hardly hybridize with Pd and Bi states~\cite{Maron&Eisenstein2000}. As a consequence, the Er-$4f$ states are not relevant to the low energy states near the Fermi energy ($E_F$). Therefore, we placed the Er-$4f$ electrons inside the core and represented all the core electrons by the projector-augmented-wave potential~\cite{Kresse&Hafner1993,Kresse&Joubert1999}. Spin-orbit coupling was included in all calculations.

The calculated bulk band structure of ErPdBi is shown in Fig.~\ref{Fig.5}. The lowest conduction and highest valence bands with $\Gamma_8$ symmetry ($j = 3/2$) are degenerate at $E_F$ at the $\Gamma$ point due to the cubic symmetry, resulting in a zero-gap semimetal. This semimetallic feature is consistent with the magnetotransport measurements (see Fig.~1). The spin-orbit coupling split-off $\Gamma_7$ state ($j  = 1/2$) is below the $\Gamma_6$ state. One can clearly see a band inversion between the $\Gamma_8$ and $\Gamma_6$ bands, where the $\Gamma_8$ bands are mainly contributed by Pd-$4d$ and Bi-$6p$ orbitals, while $\Gamma_6$ by Pd-$5s$ and Bi-$6s$ orbitals. Regardless of magnetic moments from the Er-$4f$ states, this band inversion means that (undoped) ErPdBi is a topological insulator, similar to HgTe and other Half Heusler topological insulators~\cite{Chadov2010,Lin2010,Feng2010}. Robust topological states are expected to exist on the surface. More interestingly, the magnetism from Er-$4f$ states can interplay with these topological surface states and generate exotic magnetoelectric effects~\cite{Qi2008}. Since the ErPdBi crystals are slightly $p$-doped as concluded from the Hall data (Fig.~\ref{Fig.1}), the real $E_F$ is expected to lie marginally below the $\Gamma_8$ degenerate point with a small hole-pocket, as illustrated in Fig.~\ref{Fig.5}. The bulk superconductivity can be attributed to these heavy-hole $\Gamma_8$ states.

\section{Summary} Electrical transport, ac-susceptibility and dc-magnetization measurements provide solid evidence for superconductivity at 1.22~K and antiferromagnetic order at 1.06~K in the non-centrosymmetric Half Heusler compound ErPdBi. The combination of superconductivity and AFM order is unusual. Possibly, the ordering phenomena occur in different electron subsystems: superconductivity in the low-carrier hole band and local moment magnetism due to Er $4f$-moments. However, since $T_N \simeq T_c$, and ErPdBi lacks inversion symmetry, the interplay of superconductivity and magnetism might give rise to a complex ground state. Electronic structure calculations show ErPdBi has an inverted band order and thus may harbor topological quantum states. We conclude the Half Heusler REPdBi series provides a unique opportunity to investigate the interplay of antiferromagnetic order, superconductivity and topological quantum states.

\acknowledgments{
This work is part of the research programme on Topological Insulators of the Foundation for Fundamental Research on Matter (FOM), which is part of the Netherlands Organisation for Scientific Research (NWO). B.Y. acknowledges financial support from a European Research Council (ERC) Advanced Grant (291472).}

\bibliography{RefsTI}

\end{document}